\pdfoutput=1
\documentclass[11pt]{article}

\usepackage[letterpaper,margin=1in]{geometry}
\usepackage[T1]{fontenc}
\usepackage[utf8]{inputenc}
\usepackage{lmodern}
\usepackage{amsmath}
\usepackage{amsfonts}
\usepackage{amssymb}
\usepackage{graphicx}
\usepackage{booktabs}
\usepackage{caption}
\usepackage{subcaption}
\usepackage{algorithm}
\usepackage{algpseudocode}
\usepackage{enumitem}
\usepackage{microtype}
\usepackage{xspace}
\usepackage{xcolor}
\usepackage{url}
\usepackage[colorlinks=true,linkcolor=blue!60!black,citecolor=blue!60!black,urlcolor=blue!60!black]{hyperref}
\usepackage[capitalize,nameinlink]{cleveref}

\DeclareGraphicsExtensions{.pdf,.png,.jpg}
\newcommand{\freqb}{\textsc{FreqBased}\xspace}
\newcommand{\FB}{\textsc{FreqBased}\xspace}

\begin{document}

\title{A Frequency-Based Algorithm for the Maximal Square
  Problem in Binary Matrices}

\author{Swastik Bhandari\\
  Kathmandu University\\
  \texttt{swastik3223@student.ku.edu.np}}

\date{}

\maketitle

\begin{abstract}
We present \freqb{}, a frequency-based algorithm for the maximal
square problem in binary matrices.
The algorithm maintains a vertical run-length array and a horizontal
counter to validate candidate squares, matching the standard
single-array dynamic programming algorithm in $O(mn)$ time and
$O(n)$ space while reducing the amount of work performed during
execution.
Experiments on synthetic random binary matrices reveal a clear
density-dependent performance transition: \freqb{} consistently
outperforms the dynamic programming baseline on moderate- and
high-density inputs across Intel and AMD processors, achieving peak
speedups of $1.175\times$ and $1.277\times$, respectively, while
executing up to $2.08\times$ fewer instructions.
On 180 real aerial image masks from the INRIA dataset, \freqb{} outperforms the baseline on every instance with speedups of up to $4.63\times$, confirming that the instruction-count advantage observed on synthetic inputs extends to real binary matrices with spatial structure. Hardware performance counter analysis shows that the performance
advantage arises primarily from reduced instruction count despite
lower execution efficiency, demonstrating that algorithms with
identical asymptotic complexity can exhibit substantially different
practical performance depending on input characteristics.
\end{abstract}

\section{Introduction}
\label{sec:intro}

The maximal square detection problem asks for the largest contiguous square
submatrix composed entirely of $1$s in a binary matrix. It is a fundamental
primitive with applications in computer vision, pattern recognition, spatial data
analysis, and semiconductor
manufacturing~\cite{nagamura2025wafer,wu2015wafer,orlowski1990algorithm}.

\paragraph{The standard approach.}
The classical solution is a dynamic programming recurrence
that processes each cell in $O(1)$ time, yielding an
$O(mn)$ time and $O(mn)$ space algorithm~\cite{cormen2009introduction}.
A space-optimised variant, which we refer to as
\textsc{dp\_lin\_1arr}, retains only a single auxiliary
array and a diagonal scalar, reducing space to $O(n)$
while preserving $O(mn)$ time.
Both formulations visit every cell, but the dominant
per-cell work, a three-way minimum, is applied only at
cells containing a one.
The total instruction count of \textsc{dp\_lin\_1arr}
therefore grows with density, while the cost at zero
cells remains constant.

\paragraph{Our contribution.}
We present \freqb{}, a frequency-based algorithm that
maintains a vertical run-length array and a horizontal
counter to validate candidate squares without performing
the three-way minimum at every cell.
\freqb{} runs in $O(mn)$ time and $O(n)$ space, matching
the asymptotic complexity of \textsc{dp\_lin\_1arr}, but
its total instruction count grows slowly with density, while \textsc{dp\_lin\_1arr}'s grows
proportionally with the number of ones.

\paragraph{Key results.}
\freqb{} dominates \textsc{dp\_lin\_1arr} at every tested
density $\rho \geq 0.5$ across all matrix sizes and both
platforms, with speedups reaching $1.175\times$ on Intel
and $1.277\times$ on AMD at $\rho = 0.9$, $N = 20{,}000$.
On the INRIA real-world dataset, \freqb{} wins all 180
aerial image tiles with speedups up to $4.630\times$,
including tiles well below the synthetic crossover density,
where spatial structure reduces its branch miss rate by
$24.7\times$ relative to i.i.d.\ inputs at the same density.
Hardware performance counter analysis attributes the advantage
to instruction-count reduction ($2.08\times$ fewer
instructions at $\rho = 0.9$), demonstrating that two
$O(mn)$ algorithms can differ substantially in practice
when input density and spatial structure vary.

\paragraph{Organization.}
Section~\ref{sec:related} reviews related work.
Section~\ref{sec:algorithm} presents the algorithm and its
correctness and complexity analysis.
Section~\ref{sec:results} presents experimental results.
Applications are discussed in Section~\ref{sec:applications}.


\section{Related Work}
\label{sec:related}

The $O(N^{2})$ DP solution for maximal square is standard textbook
material~\cite{cormen2009introduction}. The related maximal rectangle problem is
solvable in $O(N^{2})$ using a histogram-based sweep, which serves as the basis
for one of our evaluated baselines. Largest empty rectangle variants have been
studied by Orlowski~\cite{orlowski1990algorithm}, with extensions to
hyper-rectangles~\cite{aboalsabeh2023hyper} and mining of maximal empty
rectangles~\cite{laskar2025mining}. Sun and Nobel~\cite{sun2008submatrices}
analyse submatrix structure in random binary matrices. Jeon et
al.~\cite{jeon2009access} abstract memory-access patterns of dynamic programs,
motivating our examination of per-cell array accesses. Applications in wafer map
analysis~\cite{nagamura2025wafer,wu2015wafer} and occupancy-grid
robotics~\cite{orlowski1990algorithm,aboalsabeh2023hyper} are further described
in \cref{sec:applications}. Our experimental design follows the statistical
methodology of Amin et al.~\cite{Amin2023} and Agarwal et al.~\cite{Agarwal2021}.


\section{Algorithm Design and Analysis}
\label{sec:algorithm}

\subsection{Core Methodology}
\FB{} processes the input matrix row by row while maintaining a
one-dimensional auxiliary array $\mathit{freq}[0\,..\,\mathit{row\_size}-1]$,
where $\mathit{freq}[j]$ stores the height of the consecutive vertical run
of 1s ending at the current row in column $j$.
Each entry is incremented when the corresponding cell contains a~1 and
reset to zero otherwise, requiring $O(\mathit{row\_size})$ auxiliary
memory in addition to the input matrix.

\begin{figure}[tp]
  \centering
  \includegraphics[width=0.82\linewidth]{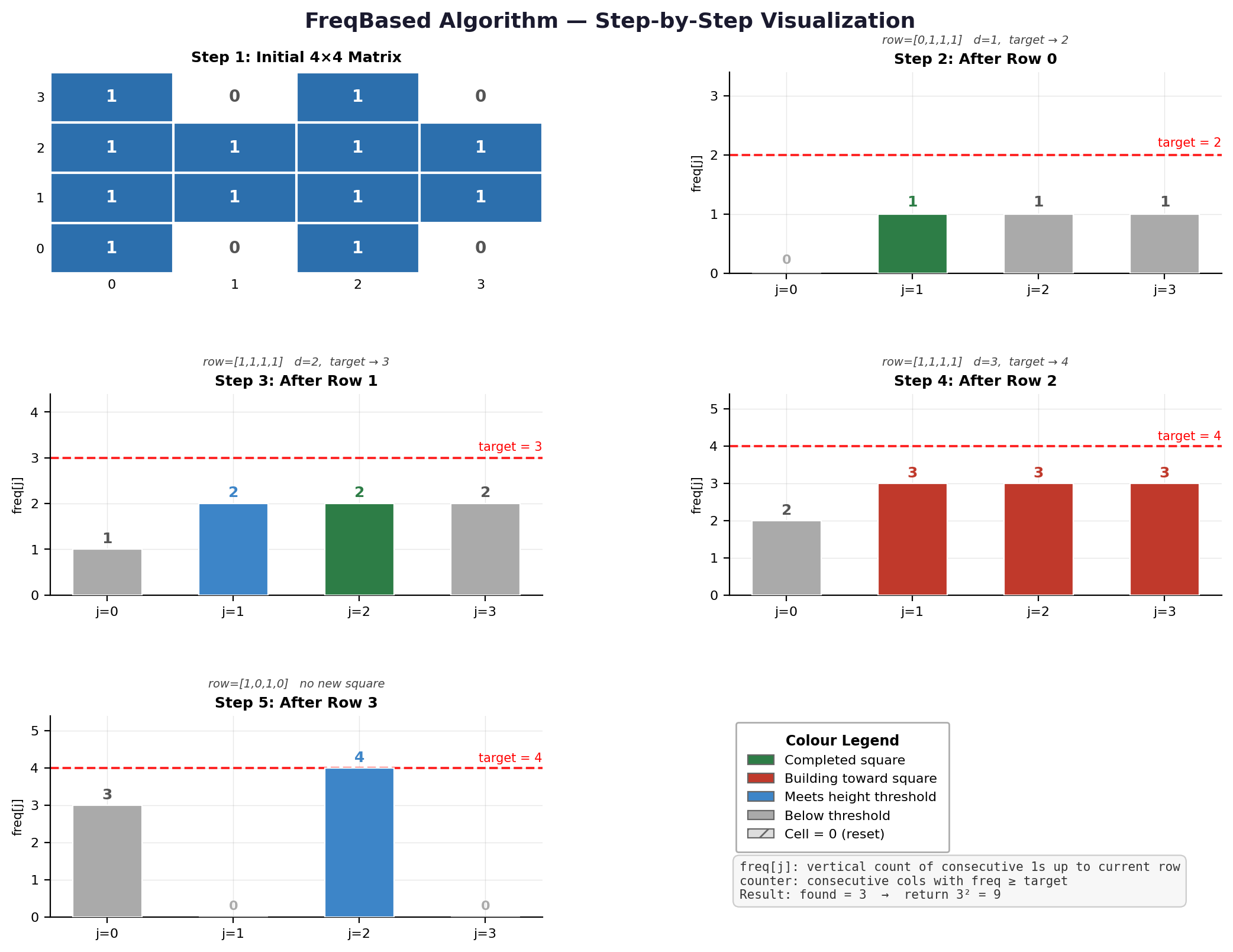}
  \caption{Step-by-step visualisation of \FB{} on a $4\times4$ binary
  matrix.}
  \label{fig:algviz}
\end{figure}

Rather than computing a candidate square size for every cell, as classical
dynamic programming formulations do, \FB{} maintains a single integer
$\mathit{check\_max\_side\_length}$, initialised to~1, representing the
smallest unconfirmed candidate side length.
The target advances by one only after a square of the current size is
confirmed, deferring evaluation of larger candidates until necessary.
Figure~\ref{fig:algviz} illustrates the evolution of the frequency array
and target side length on a small example.

\subsection{Greedy Square Validation}
At each row, columns are scanned left to right.
A column $j$ qualifies if
\[
  \mathit{freq}[j] \;\geq\; \mathit{check\_max\_side\_length};
\]
otherwise $\mathit{counter}$ resets to zero and the column is skipped.
Consecutive qualifying columns are accumulated in $\mathit{counter}$.
When $\mathit{counter}$ reaches $\mathit{check\_max\_side\_length}$
(line~\ref{ln:found-square}), a valid all-ones square of that side
length has been identified: $\mathit{found\_max\_side}$ is incremented,
$\mathit{check\_max\_side\_length}$ advances by one, and $\mathit{counter}$
resets to zero to begin searching for the next larger candidate.

\subsection{Correctness}

The algorithm relies on two invariants.
First, after processing row $i$,
$\mathit{freq}[j]$ equals the height of the maximal consecutive vertical
run of 1s ending at position $(i,j)$.
Consequently, any sequence of
$\mathit{check\_max\_side\_length}$ consecutive qualifying columns
guarantees the existence of an all-ones square of the corresponding side
length.

Second,
$\mathit{check\_max\_side\_length}$ always denotes the smallest
unverified candidate side length and it is incremented only after a
valid square of the current candidate size has been confirmed, ensuring
that each increment corresponds to the discovery of a larger feasible
square.
Therefore,
$\mathit{found\_max\_side}$ equals the maximum verified side length, and the
algorithm returns the area of the largest all-ones square.

\subsection{Complexity Analysis}

Each matrix element is visited exactly once, and updating
$\mathit{freq}$, $\mathit{counter}$, and the target side length requires
constant time per cell.
Consequently, \FB{} runs in
$O(\mathit{col\_size}\times\mathit{row\_size})$ time while requiring only
$O(\mathit{row\_size})$ auxiliary space.

\begin{algorithm}[tp]
  \caption{\FB($M$, $\mathit{col\_size}$, $\mathit{row\_size}$)}\label{alg:freqbased}
  \begin{algorithmic}[1]
    \Require Binary matrix $M\in\{0,1\}^{\mathit{col\_size}\times \mathit{row\_size}}$
    \Ensure Area of the largest all-ones square in $M$
    \State $\mathit{freq}[0..\mathit{row\_size}-1]\gets 0$
           \Comment{vertical run-length of 1s per column}
    \State $\mathit{found\_max\_side}\gets 0$;\;
           $\mathit{check\_max\_side\_length}\gets 1$;\;
           $\mathit{counter}\gets 0$
    \For{$i=0$ \textbf{to} $\mathit{col\_size}-1$}
      \State $\mathit{counter}\gets 0$
             \Comment{reset horizontal run at each new row}
      \For{$j=0$ \textbf{to} $\mathit{row\_size}-1$}
        \If{$M[i][j]=0$}
          \State $\mathit{freq}[j]\gets 0$;\;
                 $\mathit{counter}\gets 0$
          \State \textbf{continue}
        \EndIf
        \State $\mathit{freq}[j]\gets\mathit{freq}[j]+1$
               \Comment{extend vertical run of 1s}
        \If{$\mathit{freq}[j]<\mathit{check\_max\_side\_length}$}
          \State $\mathit{counter}\gets 0$
                 \Comment{insufficient height for target square}
          \State \textbf{continue}
        \EndIf
        \State $\mathit{counter}\gets\mathit{counter}+1$
        \If{$\mathit{counter}=\mathit{check\_max\_side\_length}$}
          \label{ln:found-square}
          \State $\mathit{found\_max\_side}\gets\mathit{found\_max\_side}+1$
          \State $\mathit{check\_max\_side\_length}\gets\mathit{check\_max\_side\_length}+1$
          \State $\mathit{counter}\gets 0$
                 \Comment{restart search for next larger square}
        \EndIf
      \EndFor
    \EndFor
    \State \Return $\mathit{found\_max\_side}^{2}$
  \end{algorithmic}
\end{algorithm}
 
\section{Experimental Setup}

\subsection{Hardware Platforms}

Experiments were conducted on two commodity laptop platforms running
Ubuntu~26.04~LTS and GCC~15.2.0.
Table~\ref{tab:hardware} summarizes the hardware specifications.
All implementations were compiled with \texttt{-march=native}, which
resolved to \texttt{znver3} on the AMD platform and
\texttt{skylake} on the Intel platform.

\begin{table}[htbp]
\centering
\footnotesize\setlength{\tabcolsep}{2.5pt}
\caption{Hardware platforms used for benchmarking.}
\label{tab:hardware}
\begin{tabular}{lll}
\toprule
\textbf{Property} & \textbf{AMD} & \textbf{Intel} \\
\midrule
CPU & AMD Ryzen 5 7535HS & Intel Core i5-9300H \\
Microarchitecture & \texttt{znver3} & \texttt{skylake} \\
Cores / Threads & 6 / 12 & 4 / 8 \\
L1d cache & 32\,KiB/core & 32\,KiB/core \\
L2 cache & 512\,KiB/core & 256\,KiB/core \\
L3 cache & 16\,MiB shared & 8\,MiB shared \\
RAM & 14\,GiB & 7.1\,GiB \\
OS & Ubuntu 26.04 LTS & Ubuntu 26.04 LTS \\
Compiler & GCC 15.2.0 & GCC 15.2.0 \\
\bottomrule
\end{tabular}
\end{table}

\subsection{Algorithms}

We compare four maximal-square algorithms:

\begin{itemize}
\item \textbf{FreqBased}: the frequency-based algorithm proposed in this work.
\item \textbf{dp\_lin\_1arr}: the standard dynamic-programming recurrence using a single auxiliary array.
\item \textbf{dp\_quad}: the classical dynamic-programming algorithm using a full auxiliary matrix.
\item \textbf{hist}: a histogram-based algorithm using a monotone stack.
\end{itemize}

All implementations operate on the same binary matrix representation.
Correctness was verified prior to benchmarking using 150,000 randomly
generated matrices (sizes $1$--$6$, densities $0.2$--$0.9$), with every
algorithm compared against a brute-force reference implementation.
Benchmark execution aborts if any mismatch is detected.

\subsection{Compilation and Reproducibility}

All implementations were compiled identically using

\begin{verbatim}
g++ -O2 -march=native -DNDEBUG -flto
\end{verbatim}

To reduce measurement noise, turbo boost was disabled, benchmarks were
pinned to a dedicated CPU core using
\texttt{sched\_setaffinity}, and Linux performance counters were enabled
through \texttt{perf\_event\_paranoid=1}. The same settings were used on
both platforms.

\subsection{Timing and Performance Counters}

Wall-clock time was measured using
\texttt{clock\_gettime(CLOCK\_MONOTONIC)}.
For each benchmark configuration, the repetition count was
automatically increased until the cumulative execution time exceeded
200\,ms (or $2^{20}$ repetitions).
Seven independent timing trials were then collected, and the median
per-invocation runtime was reported.
The coefficient of variation (CV) across the seven trials was recorded
to quantify measurement stability.

Hardware performance counters were collected using
\texttt{perf\_event\_open} in user-space-only mode.
The following events were measured:

\begin{itemize}
\item retired instructions,
\item CPU cycles,
\item retired branch instructions,
\item branch mispredictions,
\item L1 data-cache read misses.
\end{itemize}

Counter values were corrected using the kernel's
\texttt{time\_enabled/time\_running} scaling factor and normalized
per benchmark invocation.
From these measurements we derive instructions per cycle (IPC), branch
mispredictions per thousand instructions (bmpki), and L1 data-cache
misses per thousand instructions (l1\_mpki).

\subsection{Synthetic Benchmark}

Synthetic inputs are square binary matrices generated by independently
setting each cell to one with probability $\rho$.
Matrices were generated using \texttt{std::mt19937} with fixed seeds,
ensuring identical workloads on both platforms.

\begin{table}[tb]
\centering
\footnotesize
\caption{Synthetic benchmark parameter space.}
\label{tab:synthetic-params}
\begin{tabular}{ll}
\toprule
Parameter & Values \\
\midrule
Matrix size ($N$) & 2500, 5000, 10000, 15000, 20000 \\
Density ($\rho$) & 0.1, 0.3, 0.5, 0.7, 0.9 \\
Independent seeds & 5 \\
Algorithms & 4 \\
Runs per platform & 500 \\
\bottomrule
\end{tabular}
\end{table}

Results for each configuration were averaged across the five independent
seeds.

\subsection{Real-World Benchmark}

To complement the synthetic workload, we evaluated all algorithms on the
INRIA Aerial Image Labeling dataset~\cite{maggiori2017inria}, using the
180 publicly available training masks.
Each binary mask has resolution $5000 \times 5000$ and was used without
resizing or geometric preprocessing.
Pixels with intensity greater than 127 were mapped to foreground.

Unlike the synthetic workload, INRIA masks exhibit strong spatial
correlation due to building footprints and urban layouts, allowing us to
evaluate algorithm behavior on realistic binary images.

Table~\ref{tab:inria-density} summarizes the foreground density of each
city.

\begin{table}[tb]
\centering
\footnotesize
\caption{Per-city foreground density in the INRIA training set.}
\label{tab:inria-density}
\begin{tabular}{lrrrrr}
\toprule
City & Tiles & Mean & Median & Min & Max \\
\midrule
Austin   & 36 & 0.150 & 0.156 & 0.074 & 0.237 \\
Chicago  & 36 & 0.243 & 0.234 & 0.144 & 0.405 \\
Kitsap   & 36 & 0.049 & 0.029 & 0.002 & 0.164 \\
Tyrol-w  & 36 & 0.058 & 0.048 & 0.010 & 0.179 \\
Vienna   & 36 & 0.288 & 0.291 & 0.071 & 0.488 \\
\midrule
All & 180 & 0.158 & 0.145 & 0.002 & 0.488 \\
\bottomrule
\end{tabular}
\end{table}

The entire dataset lies within the low-to-moderate density regime
($\rho < 0.5$), complementing the broader synthetic density sweep
($0.1 \leq \rho \leq 0.9$).
Real-data experiments were conducted on the AMD platform using the same
timing and measurement methodology as the synthetic benchmarks.

\section{Experimental Results}
\label{sec:results}

\subsection{Competitive Landscape}
\label{subsec:landscape}
\begin{figure}[tp]
  \centering
  \begin{subfigure}[b]{0.48\linewidth}
    \includegraphics[width=\linewidth]{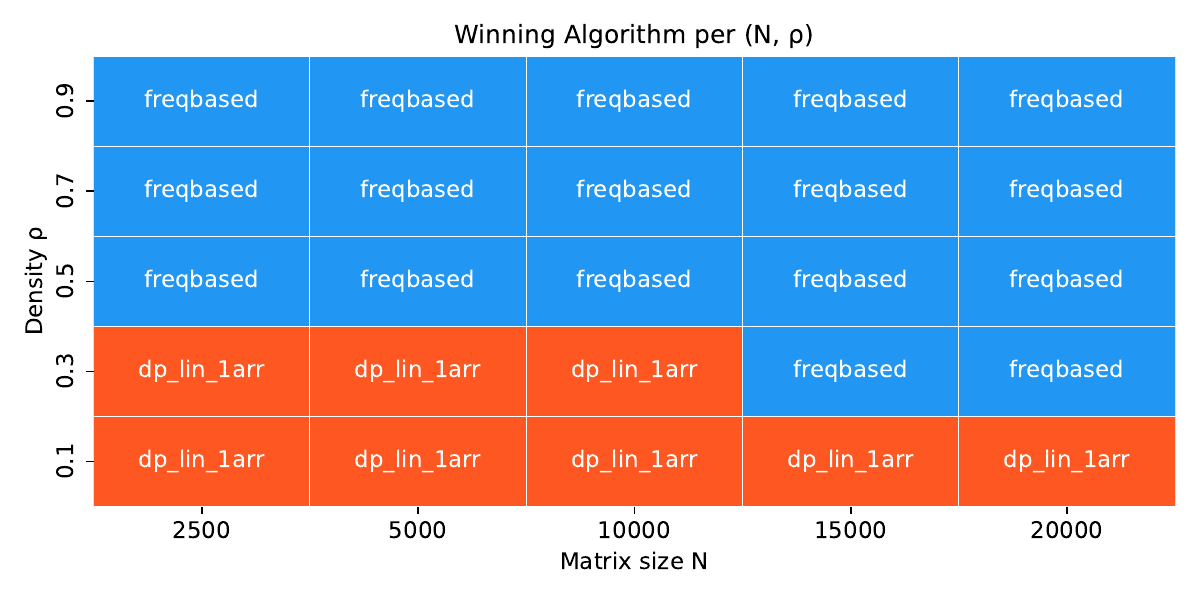}
    \caption{Intel Core i5-9300H (\texttt{skylake})}
  \end{subfigure}
  \hfill
  \begin{subfigure}[b]{0.48\linewidth}
    \includegraphics[width=\linewidth]{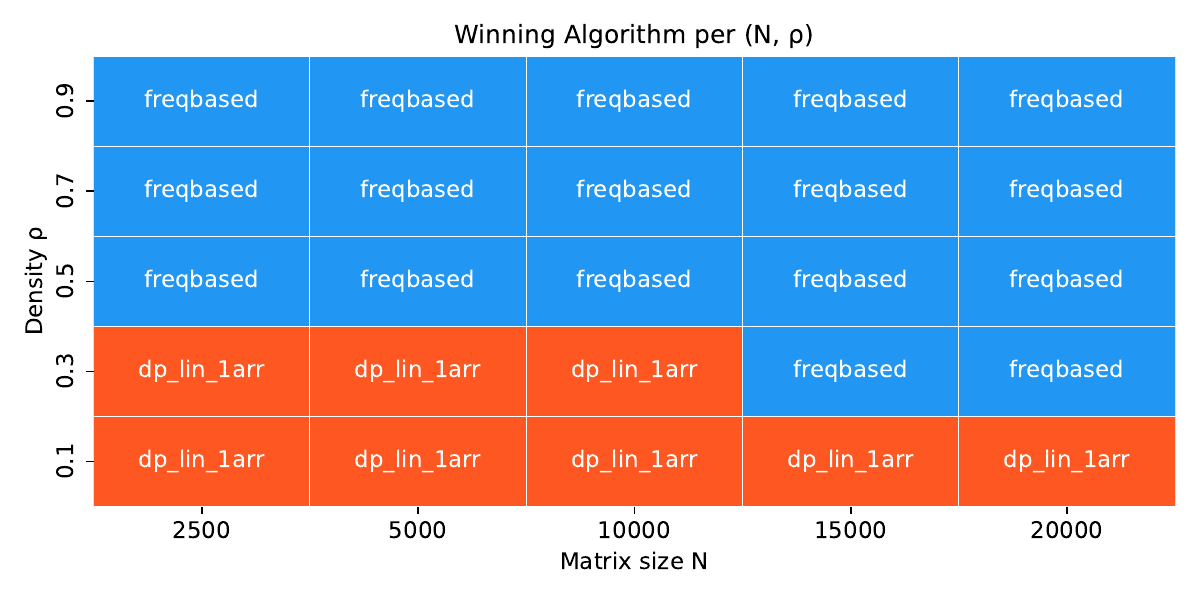}
    \caption{AMD Ryzen~5 7535HS (\texttt{znver3})}
  \end{subfigure}
  \caption{Winning algorithm at each $(N,\rho)$ configuration.
           Blue indicates \freqb{}; orange indicates
           \textsc{dp\_lin\_1arr}.}
  \label{fig:winner-pivot}
\end{figure}

Across all 25 $(N, \rho)$ configurations on both platforms,
\textsc{dp\_quad} and \textsc{hist} win zero configurations and
are strictly dominated by \textsc{dp\_lin\_1arr} at every tested
point.
Both are excluded from further analysis.
The two competitive algorithms are \freqb{} (17 wins) and
\textsc{dp\_lin\_1arr} (8 wins).
Figure~\ref{fig:winner-pivot} shows which algorithm achieves the
minimum wall-clock time at each $(N,\rho)$ pair; the pattern is
identical on both the AMD and Intel platforms.
A sharp regime boundary separates the two algorithms:
\textsc{dp\_lin\_1arr} wins at low density ($\rho \leq 0.1$
for all $N$; $\rho = 0.3$ for $N \leq 10{,}000$), while
\freqb{} wins at moderate to high density ($\rho \geq 0.5$
for all $N$, and $\rho \geq 0.3$ for $N \geq 15{,}000$).
Because \textsc{dp\_lin\_1arr} is the only algorithm that ever
defeats \freqb{}, all subsequent analysis uses it as the sole
baseline.
Speedup is defined throughout as
$t_{\textsc{dp\_lin\_1arr}} / t_{\freqb{}}$, so values
greater than 1.0 indicate \freqb{} is faster.

\subsection{Density-Driven Phase Transition}
\label{subsec:phase}

\begin{figure}[tp]
  \centering
  \begin{subfigure}[b]{0.48\linewidth}
    \includegraphics[width=\linewidth]{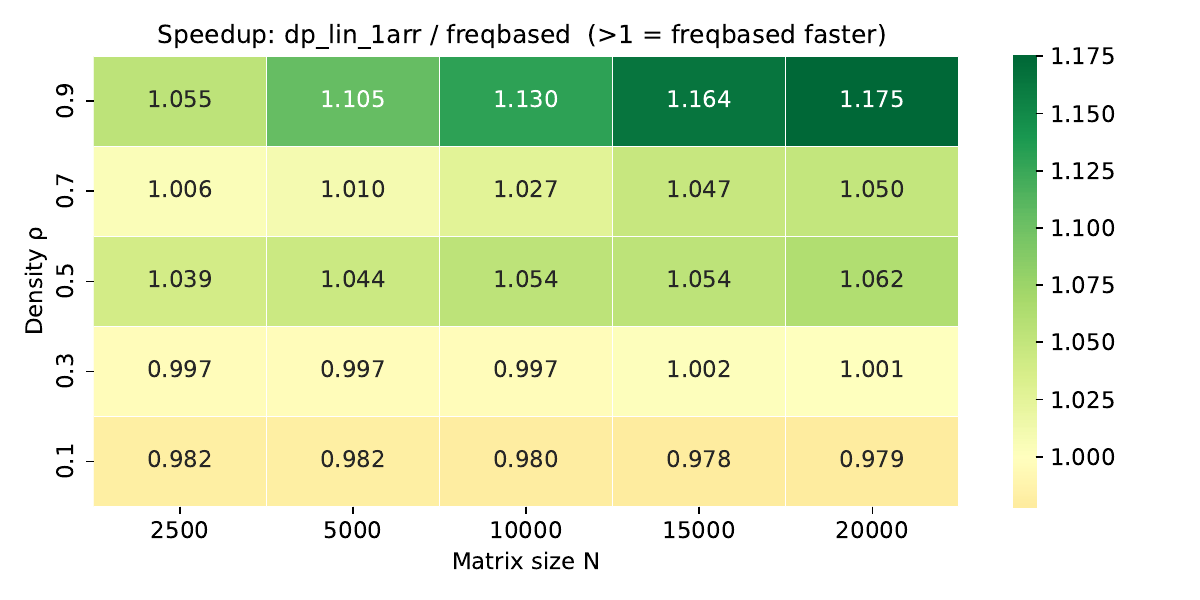}
    \caption{Intel Core i5-9300H (\texttt{skylake})}
  \end{subfigure}
  \hfill
  \begin{subfigure}[b]{0.48\linewidth}
    \includegraphics[width=\linewidth]{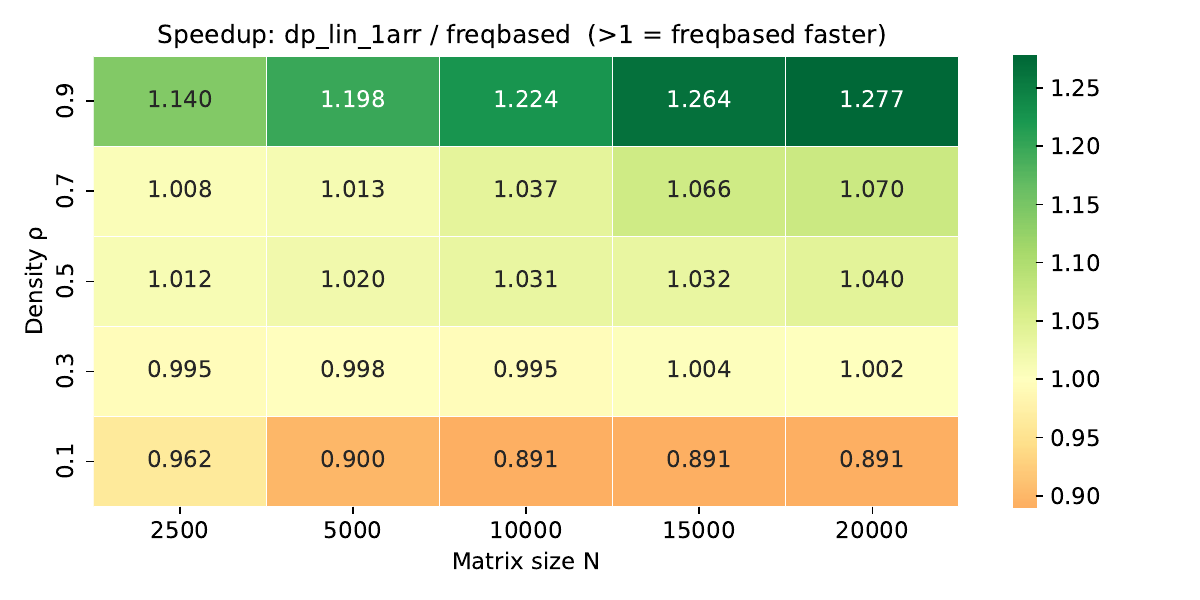}
    \caption{AMD Ryzen~5 7535HS (\texttt{znver3})}
  \end{subfigure}
  \caption{Speedup of \freqb{} over \textsc{dp\_lin\_1arr}
           ($t_{\textsc{dp}} / t_{\freqb{}}$).
           Values above $1.0$ indicate \freqb{} is faster.
           }
  \label{fig:speedup-heatmap}
\end{figure}

Figure~\ref{fig:speedup-heatmap} shows the speedup of \freqb{} over
\textsc{dp\_lin\_1arr} across all $(N, \rho)$ pairs on the Intel
platform, with AMD shown for cross-architecture comparison.
The dominant structure is horizontal: the regime boundary runs along
density, not matrix size, confirming that $\rho$ is the primary
determinant of which algorithm wins.

Three observations follow directly from the data.

\paragraph{The crossover occurs between $\rho=0.3$ and $\rho=0.5$.}
At $\rho \leq 0.1$, \textsc{dp\_lin\_1arr} wins across all five
matrix sizes.
At $\rho \geq 0.5$, \freqb{} wins across all five matrix sizes.
At $\rho = 0.3$, the two algorithms are essentially at parity for
$N \leq 10{,}000$; \freqb{} becomes the faster choice only for
$N \geq 15{,}000$.

\paragraph{Losses at low density are bounded.}
Losses at low density are bounded. At $\rho = 0.1$, \freqb{} is
slower by at most $2.2\%$ on Intel and $10.9\%$ on AMD across all
tested matrix sizes.

\paragraph{The regime structure is consistent across architectures.}
The winner at every $(N, \rho)$ pair is identical on Intel and AMD
(Figure~\ref{fig:winner-pivot}).
Speedup magnitudes differ quantitatively: \freqb{} peaks at
$1.175\times$ on Intel and $1.277\times$ on AMD, both at
$\rho=0.9$, $N=20{,}000$ (Figure~\ref{fig:speedup-heatmap}).

\subsection{Speedup Scaling with $N$}
\label{subsec:speedup}

\begin{figure}[tp]
  \centering
  \begin{subfigure}[b]{0.48\linewidth}
    \includegraphics[width=\linewidth]{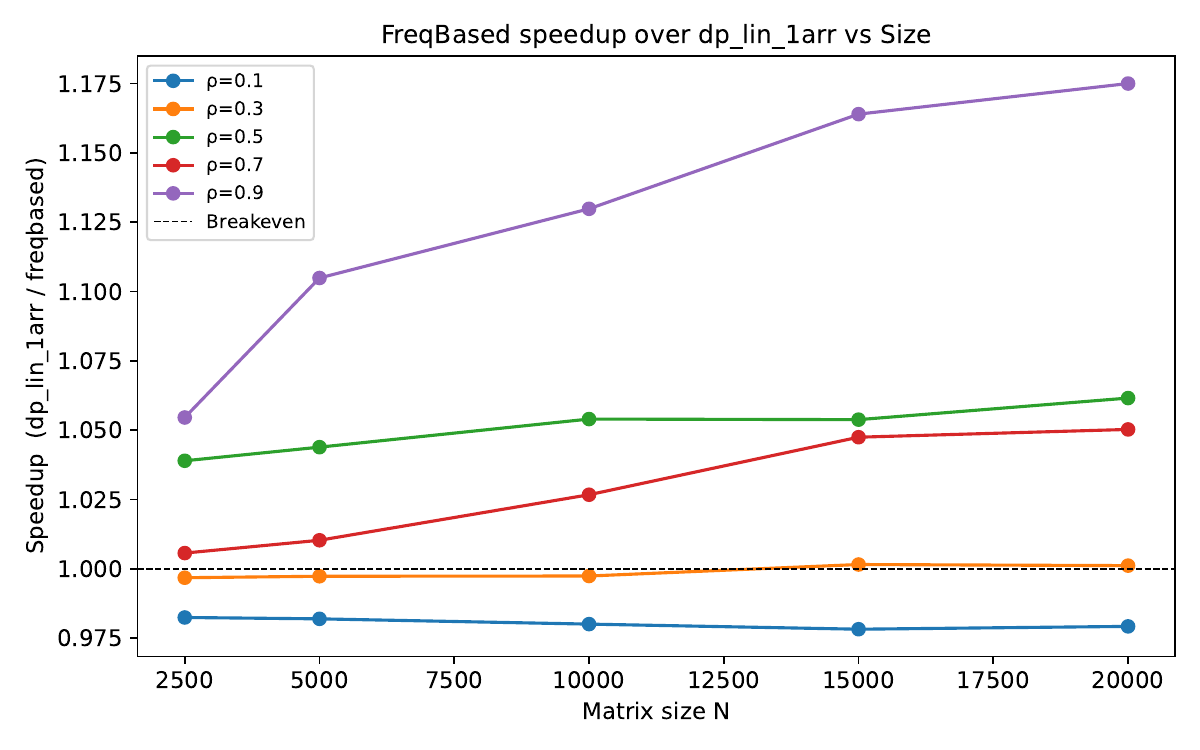}
    \caption{Intel Core i5-9300H (\texttt{skylake})}
  \end{subfigure}
  \hfill
  \begin{subfigure}[b]{0.48\linewidth}
    \includegraphics[width=\linewidth]{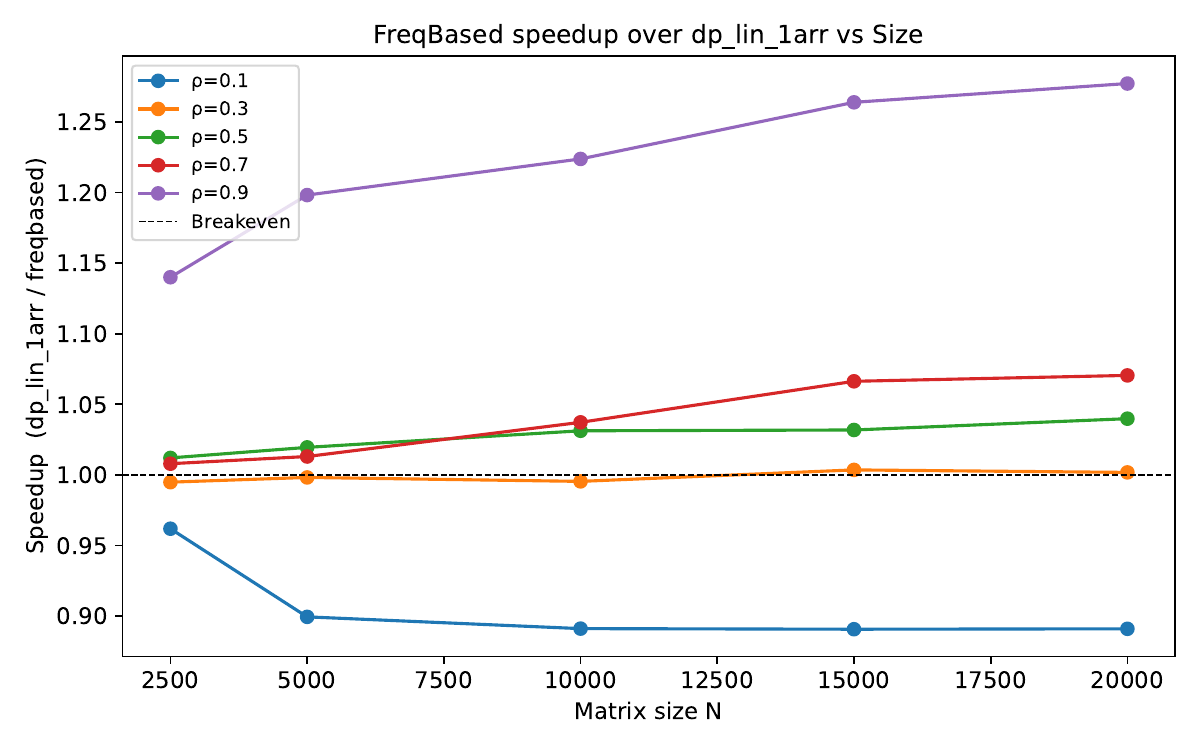}
    \caption{AMD Ryzen~5 7535HS (\texttt{znver3})}
  \end{subfigure}
  \caption{Speedup of \freqb{} over \textsc{dp\_lin\_1arr} as a
           function of matrix size $N$, for each density level.
           The dashed line marks breakeven ($1.0\times$).
           }
  \label{fig:speedup-vs-N}
\end{figure}

Figure~\ref{fig:speedup-vs-N} shows how speedup varies with matrix
size $N$ at each density level on Intel, with AMD shown for
comparison.
Two patterns are immediately visible.

\paragraph{At high density, speedup grows with $N$.}
At $\rho = 0.9$, speedup increases from $1.055\times$ at
$N = 2{,}500$ to $1.175\times$ at $N = 20{,}000$, and has not
saturated within the tested range.
At $\rho = 0.7$, the same trend holds, growing from $1.006\times$
to $1.050\times$.
Both values are directly readable from
Figure~\ref{fig:speedup-vs-N}(a).

\paragraph{At low density, speedup is stable across $N$.}
At $\rho = 0.1$, speedup remains flat near $0.98\times$ on Intel
across all five matrix sizes. On AMD, speedup drops from $0.962\times$
at $N = 2{,}500$ to $0.891\times$ at $N = 5{,}000$ before
stabilising, suggesting a size-dependent effect at low density that
does not appear on Intel.

\subsection{Work Reduction vs.\ Hardware Efficiency}
\label{subsec:causal}

\begin{figure}[tp]
  \centering
    \includegraphics[width=\linewidth]{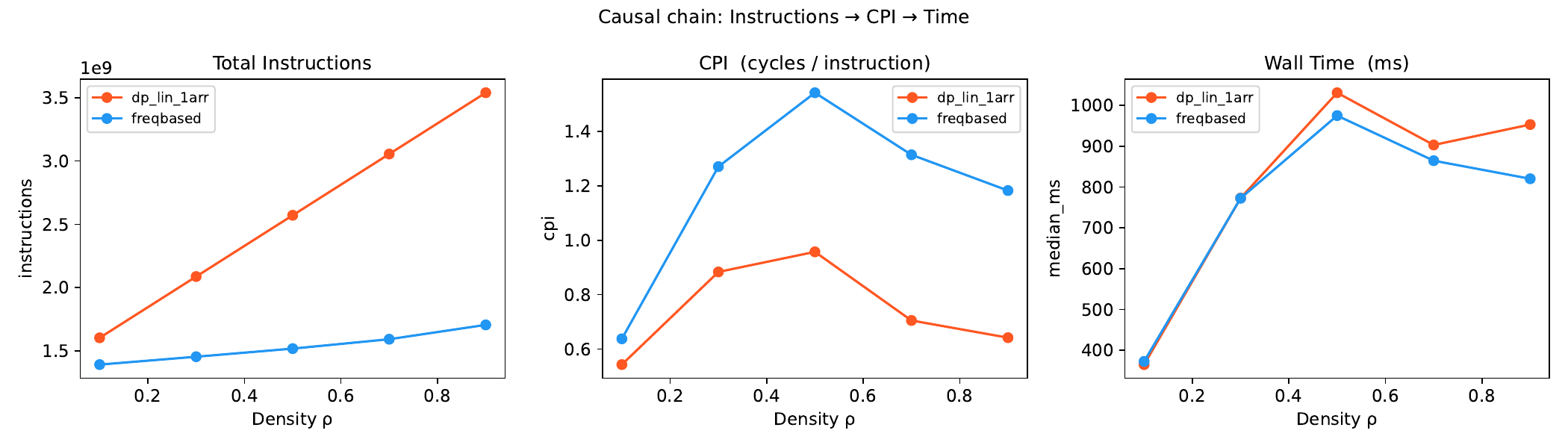}
    
    \caption{Causal decomposition on Intel Core i5-9300H: total instructions, CPI, and
           wall time versus density $\rho$.}
    \label{fig:causal-chain}
  \end{figure}

Figure~\ref{fig:causal-chain} decomposes the performance difference
into three quantities measured on Intel: total retired instructions,
cycles per instruction (CPI), and wall time. The AMD platform shows the same qualitative pattern
(Figure~\ref{fig:amd-causal-chain}).
Together they reveal why \freqb{} wins at high density despite being
a less efficient executor of each instruction.

\begin{figure}[tp]
  \centering
  \begin{subfigure}[b]{0.48\linewidth}
    \includegraphics[width=\linewidth]{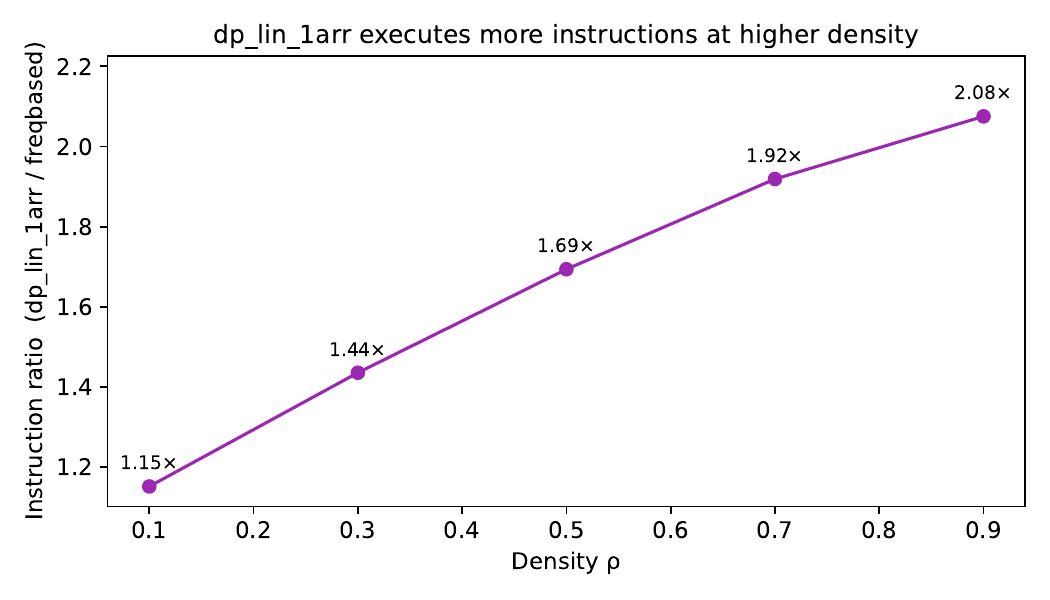}
    \caption{Intel Core i5-9300H (\texttt{skylake})}
  \end{subfigure}
  \hfill
  \begin{subfigure}[b]{0.48\linewidth}
    \includegraphics[width=\linewidth]{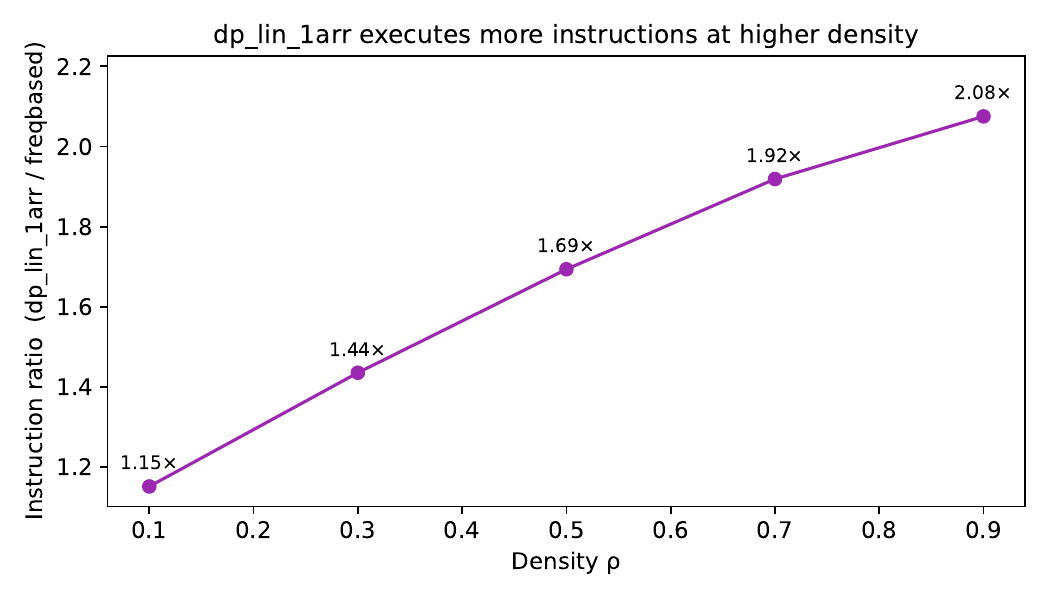}
    \caption{AMD Ryzen~5 7535HS (\texttt{znver3})}
  \end{subfigure}
  \caption{Instruction count ratio
           ($\text{instructions}_{\textsc{dp}} \;/\;
           \text{instructions}_{\freqb{}}$) versus density $\rho$.}
  \label{fig:instr-ratio}
\end{figure}

\paragraph{Instruction count.}
Figure~\ref{fig:instr-ratio} shows the ratio of instructions
executed by \textsc{dp\_lin\_1arr} to those executed by \freqb{},
averaged over all $N$.
The ratio grows monotonically with density: $1.15\times$ at
$\rho=0.1$, $1.44\times$ at $\rho=0.3$, $1.69\times$ at $\rho=0.5$,
$1.92\times$ at $\rho=0.7$, and $2.08\times$ at $\rho=0.9$.
The instruction count of \freqb{} grows modestly across densities
(Figure~\ref{fig:causal-chain}, left panel), while
\textsc{dp\_lin\_1arr} grows steeply.
\freqb{} reduces total work, and the reduction grows with density.

\paragraph{CPI.}
Despite executing fewer instructions, \freqb{} incurs a higher CPI
at every tested $(N, \rho)$ configuration, with zero exceptions
(Figure~\ref{fig:causal-chain}, center panel).
\freqb{} is consistently stall-bound; we trace the source of this
penalty to branch mispredictions in Section~\ref{subsec:counters}.

\paragraph{Wall time crossover.}
Wall time is the product of instruction count and CPI divided by
clock frequency.
\freqb{} has fewer instructions but higher CPI; the outcome depends
on which effect dominates.
The two wall-time curves converge near $\rho=0.3$ and diverge in
favour of \freqb{} at higher densities, as visible in
Figure~\ref{fig:causal-chain}, right panel.
\freqb{} wins not by executing instructions more efficiently, but
by executing fewer of them.

\subsection{Microarchitectural Profile}
\label{subsec:counters}

\begin{figure}[tp]
  \centering
  \includegraphics[width=\linewidth]{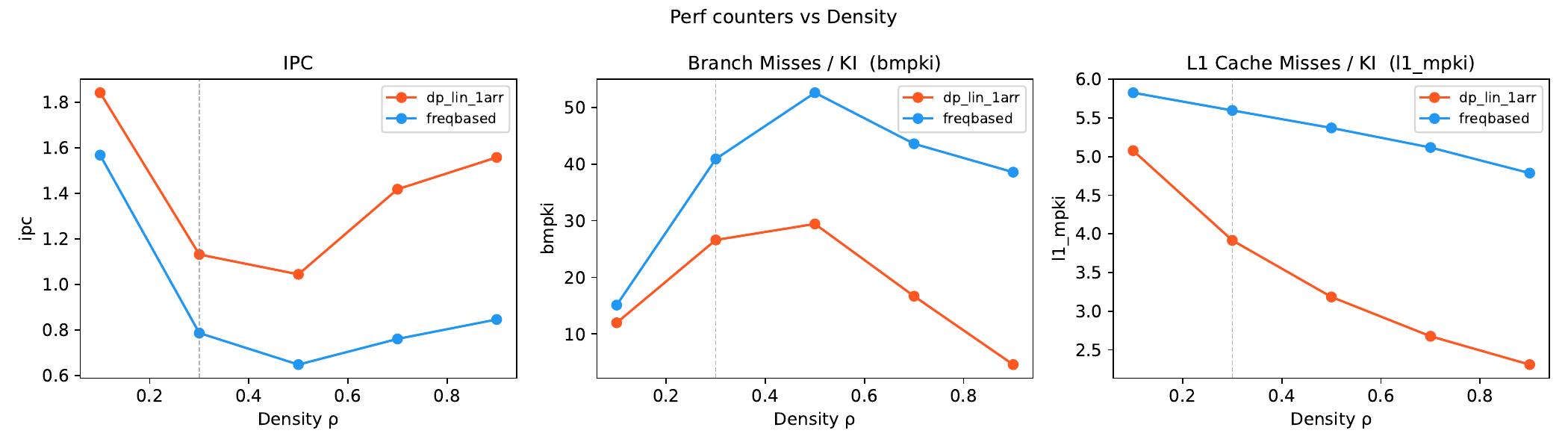}
  \caption{IPC, bmpki, and l1\_mpki versus density $\rho$
           on Intel Core i5-9300H (\texttt{skylake}).}
  \label{fig:counters-density}
\end{figure}

\begin{figure}[tp]
  \centering
  \includegraphics[width=\linewidth]{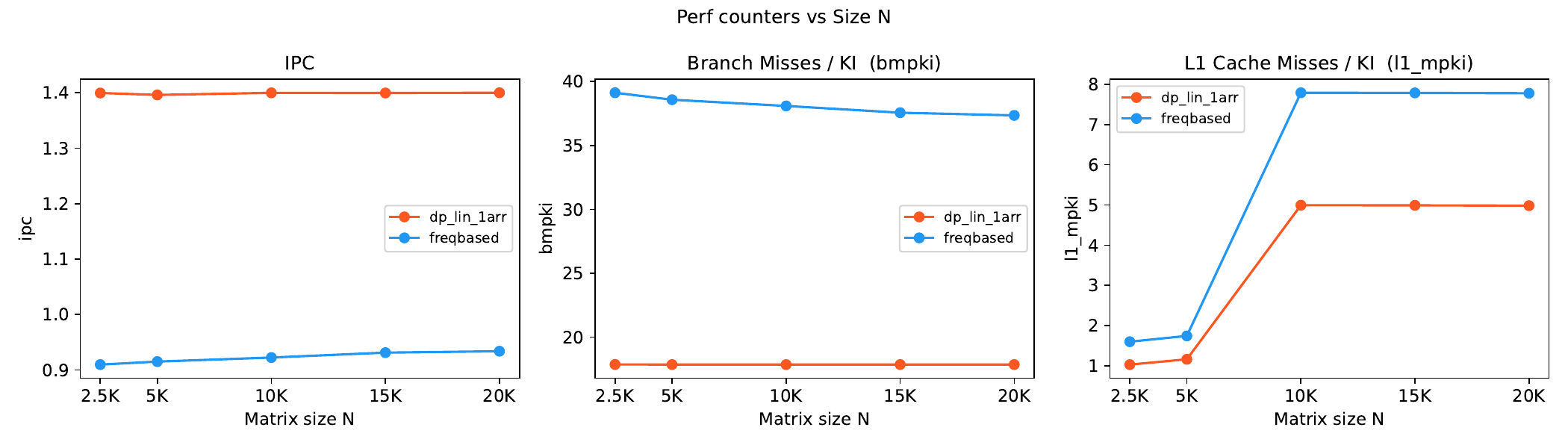}
  \caption{IPC, bmpki, and l1\_mpki versus matrix size $N$
           on Intel Core i5-9300H (\texttt{skylake}).}
  \label{fig:counters-N}
\end{figure}

To understand the execution-efficiency tradeoff identified in
Section~\ref{subsec:causal}, we examine IPC, branch mispredictions,
and L1 cache behaviour.
Figure~\ref{fig:counters-density} shows IPC, branch mispredictions
per thousand instructions (bmpki), and L1 cache read misses per
thousand instructions (l1\_mpki) versus density on Intel.
Figure~\ref{fig:counters-N} shows the same metrics versus matrix
size $N$.
The AMD platform shows the same qualitative patterns
(Figures~\ref{fig:amd-counters-density} and~\ref{fig:amd-counters-N}).

\paragraph{IPC and branch misses explain the CPI penalty.}
\freqb{} has strictly lower IPC than \textsc{dp\_lin\_1arr} at
every tested $(N, \rho)$ pair, with zero exceptions
(Figure~\ref{fig:counters-density}, left panel).
\freqb{} also has strictly higher bmpki than
\textsc{dp\_lin\_1arr} at every tested $(N, \rho)$ pair, again
with zero exceptions
(Figure~\ref{fig:counters-density}, center panel).
These observations are consistent with the CPI penalty identified
in Section~\ref{subsec:causal}: \freqb{} has both lower IPC and
higher branch miss rate than \textsc{dp\_lin\_1arr} at every
tested configuration.

\begin{table}[htbp]
\centering
\footnotesize
\caption{Branch miss rate and branch mispredictions per thousand
         instructions (bmpki) for \freqb{} on synthetic i.i.d.\
         inputs, by density.}
\label{tab:synthetic-miss-rate}
\begin{tabular}{lrr}
\toprule
Density ($\rho$) & Miss rate & bmpki \\
\midrule
0.1 & 3.45\% & 11.26 \\
0.3 & 11.02\% & 34.48 \\
0.5 & 17.35\% & 52.19 \\
0.7 & 13.01\% & 37.89 \\
0.9 & 11.50\% & 33.09 \\
\bottomrule
\end{tabular}
\end{table}

\paragraph{L1 cache miss rate.}
\freqb{} has strictly higher l1\_mpki than
\textsc{dp\_lin\_1arr} at every tested $(N, \rho)$ pair
(Figure~\ref{fig:counters-density}, right panel).
As density increases, l1\_mpki decreases for
\textsc{dp\_lin\_1arr} but remains relatively stable for \freqb{}.

\paragraph{Density governs micro-architectural behaviour, not $N$.}
IPC and bmpki are essentially stable across all five matrix sizes
for both algorithms (Figure~\ref{fig:counters-N}, left and center
panels).
The character of the computation is determined by density, not
by problem scale.
The sole exception is l1\_mpki, which increases sharply
between $N=5{,}000$ and $N=10{,}000$ for both algorithms across
every density level, then plateaus through $N=20{,}000$
(Figure~\ref{fig:counters-N}, right panel).
We report this as an observed inflection and do not attribute
a cause without further evidence.

\subsection{Real-World Validation: INRIA Aerial Image Dataset}
\label{subsec:inria}
Notably, the INRIA tiles span densities well below $\rho = 0.3$,
including entire cities with mean foreground density $\rho < 0.06$,
where \textsc{dp\_lin\_1arr} is faster on i.i.d.\ synthetic inputs.
This suggests that density alone does not determine performance on
structured real-world data.
We hypothesize that the geometric structure of building footprints leads to more predictable execution behavior for \freqb{} than independent random inputs at the same density. This hypothesis is evaluated through the hardware performance counter analysis presented later in this section.

We evaluate all four algorithms on the 180 ground-truth building
footprint masks of the INRIA Aerial Image Labeling training
split~\cite{maggiori2017inria}, each a $5000 \times 5000$ binary
matrix.
Tile densities range from $0.002$ to $0.488$; no tile exceeds
$\rho = 0.49$, placing the entire dataset within the
low-to-moderate density regime.
Measurement stability on INRIA is high: mean CV of $0.003$ and
maximum CV of $0.039$ across all 180 tiles and 4 algorithms.

\begin{figure}[htbp]
  \centering
  \includegraphics[width=0.62\linewidth]{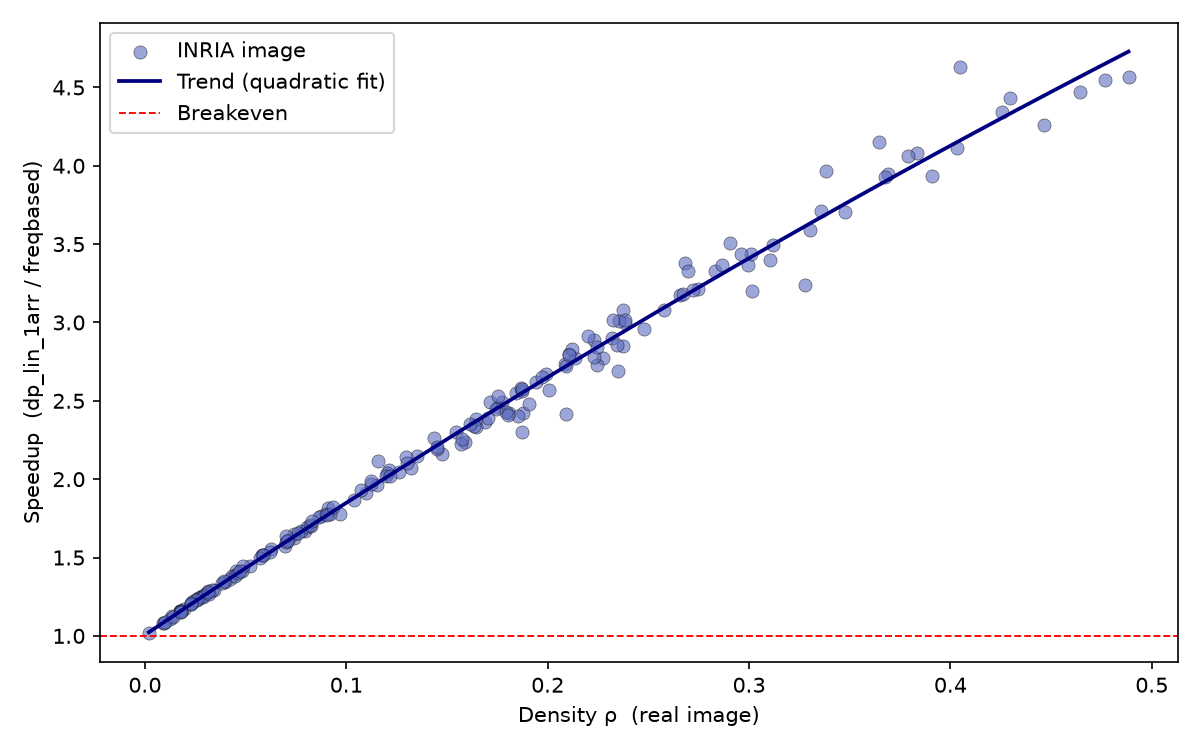}
  \caption{Speedup of \freqb{} over \textsc{dp\_lin\_1arr} versus
           tile density on 180 INRIA tiles.}
  \label{fig:real-speedup}
\end{figure}

\paragraph{\freqb{} wins on every tile.}
\freqb{} achieves the minimum wall-clock time on all 180 tiles,
with zero exceptions across all five cities
(Figure~\ref{fig:real-speedup}).
No other algorithm wins any tile.
Speedup ranges from $1.017\times$ at the lowest-density tile to
$4.630\times$ at the highest, with a strong positive relationship
between speedup and tile density ($r = 0.995$, computed over all
180 tiles).

\begin{figure}[htbp]
  \centering
  \includegraphics[width=0.62\linewidth]{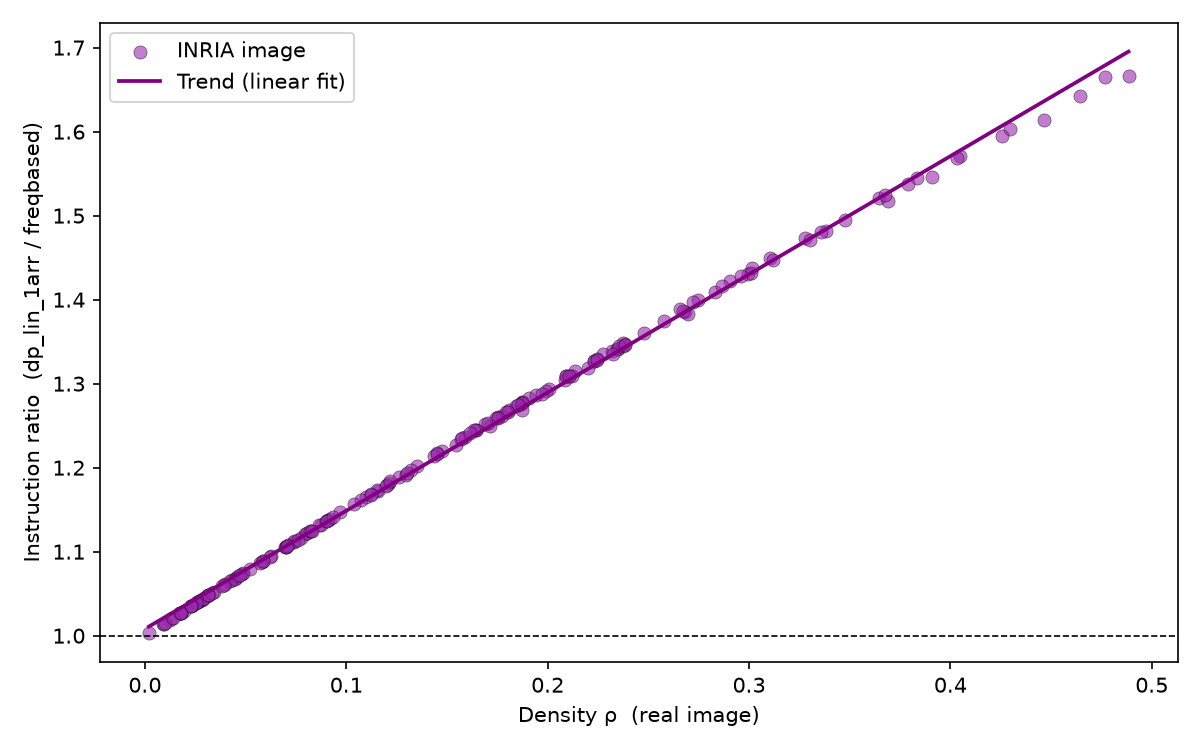}
  \caption{Instruction count ratio versus tile density on 180
           INRIA tiles.}
  \label{fig:real-instr-ratio}
\end{figure}

\paragraph{Instruction count reduction transfers to real data.}
Figure~\ref{fig:real-instr-ratio} shows the instruction count ratio
versus tile density.
The ratio exceeds $1.0$ on all 180 tiles and increases with density
($r = 0.999$, computed over all 180 tiles), confirming that the
work-reduction mechanism identified on synthetic inputs operates
identically on real binary matrices.

\paragraph{Per-city results reflect urban density.}
Table~\ref{tab:inria-city} reports per-city speedup statistics.
\freqb{} wins all 36 tiles in every city.
Mean speedup tracks mean city density directly: sparse regions
(Kitsap, mean $\rho = 0.049$) yield a mean speedup of $1.42\times$,
while dense urban areas (Vienna, mean $\rho = 0.288$) yield
$3.30\times$.
This ordering across five geographically and architecturally
distinct cities is consistent with the density-driven regime
boundary established on synthetic data in
Section~\ref{subsec:phase}.

\paragraph{Spatial structure changes the execution profile.}
Table~\ref{tab:inria-counters} shows the cycles per cell and
branch miss rate for both algorithms across density bins.
On INRIA, \freqb{} operates at a near-constant cost of
$2.0$--$2.3$ cycles per matrix cell regardless of tile density
(CV~$4.7\%$ across bins), while \textsc{dp\_lin\_1arr} climbs
from $2.5$ to $7.7$ cycles per cell as density increases.
Unlike on synthetic inputs, where \freqb{} has a strictly higher
branch miss rate than \textsc{dp\_lin\_1arr} at every
configuration (Section~\ref{subsec:counters}), on INRIA
\freqb{} has a lower miss rate at every density bin.
The geometric structure of building footprint masks produces a more
predictable access pattern for \freqb{}'s data-dependent branches
than independent random inputs at the same density, reducing its
branch miss rate to $0.14\%$ in the $0.05$--$0.10$ density range
versus $3.45\%$ on synthetic inputs at the same density
(Table~\ref{tab:synthetic-miss-rate}), a $24.7\times$ reduction
(Table~\ref{tab:inria-counters}).

\begin{table}[htbp]
\centering
\footnotesize
\caption{Per-city speedup of \freqb{} over \textsc{dp\_lin\_1arr}
         on INRIA (36 tiles per city)}
\label{tab:inria-city}
\begin{tabular}{lrrrr}
\toprule
City     & Mean $\rho$ & Mean speedup  & Min           & Max           \\
\midrule
Kitsap   & 0.049       & $1.42\times$  & $1.02\times$  & $2.34\times$  \\
Tyrol-w  & 0.058       & $1.50\times$  & $1.09\times$  & $2.43\times$  \\
Austin   & 0.150       & $2.21\times$  & $1.62\times$  & $2.90\times$  \\
Chicago  & 0.243       & $3.01\times$  & $2.27\times$  & $4.63\times$  \\
Vienna   & 0.288       & $3.30\times$  & $1.61\times$  & $4.57\times$  \\
\bottomrule
\end{tabular}
\end{table}

\begin{table}[htbp]
\centering
\footnotesize\setlength{\tabcolsep}{3.5pt}
\caption{Cycles per matrix cell and branch miss rate on INRIA,
         by density bin. Values averaged over tiles in each bin.}
\label{tab:inria-counters}
\begin{tabular}{lrrrr}
\toprule
& \multicolumn{2}{c}{Cycles / cell}
& \multicolumn{2}{c}{Branch miss rate} \\
\cmidrule(lr){2-3} \cmidrule(lr){4-5}
Density bin
  & \textsc{dp\_lin} & \freqb{}
  & \textsc{dp\_lin} & \freqb{} \\
\midrule
$(0.00,\,0.05]$ & 2.546 & 2.046 & 0.09\% & 0.06\% \\
$(0.05,\,0.10]$ & 3.478 & 2.109 & 0.20\% & 0.14\% \\
$(0.10,\,0.20]$ & 5.014 & 2.211 & 0.35\% & 0.25\% \\
$(0.20,\,0.50]$ & 7.679 & 2.275 & 0.41\% & 0.34\% \\
\bottomrule
\end{tabular}
\end{table}

\section{Applications}
\label{sec:applications}

\paragraph{Largest defect-free square in wafer maps.}
Wafer maps are binary matrices of defective and non-defective
dies~\cite{nagamura2025wafer,wu2015wafer}. Encoding non-defective dies as~$1$,
the largest all-$1$ square submatrix identifies the largest contiguous
defect-free square region.

\paragraph{Largest free square in occupancy grids.}
In binary occupancy grids, the largest free (all-zero) square gives the largest
collision-free bounding region for path planning. Inverting the matrix reduces
this to the standard maximal square
task~\cite{orlowski1990algorithm,aboalsabeh2023hyper}.


\section{Conclusion}
\label{sec:conclusion}

We presented \freqb{}, a frequency-based algorithm for the maximal
square problem that exploits vertical run-length structure to reduce
total instruction count relative to the standard single-array DP
baseline.

On synthetic i.i.d.\ binary matrices, \freqb{} wins 17 of 25
tested $(N, \rho)$ configurations on both an Intel Core i5-9300H
and an AMD Ryzen~5 7535HS, with a peak speedup of $1.175\times$
on Intel and $1.277\times$ on AMD at $\rho = 0.9$, $N = 20{,}000$.
The advantage is mechanistic: at $\rho = 0.9$, \freqb{} executes
$2.08\times$ fewer instructions than \textsc{dp\_lin\_1arr},
and the reduction grows monotonically with density.
The cost is a higher branch miss rate and lower IPC at every
tested configuration; \freqb{} wins not by executing instructions
more efficiently but by executing fewer of them.
On the INRIA Aerial Image Labeling dataset, \freqb{} wins all
180 tiles with speedups ranging from $1.017\times$ to
$4.630\times$, and the branch miss rate on real building
footprint masks is $24.7\times$ lower than on synthetic inputs
at the same density. This reduction in branch miss rate is consistent with the improved performance observed on the real dataset.

Based on the configurations tested, \freqb{} is the faster
choice at moderate to high density ($\rho \geq 0.5$ on
i.i.d.\ inputs, or lower densities when the input exhibits
strong spatial structure), while \textsc{dp\_lin\_1arr}
is preferred for sparse i.i.d.\ inputs with $\rho \leq 0.1$.    
At intermediate densities, the two algorithms perform within
$0.4\%$ of each other and the choice has negligible practical
impact.

The primary limitation of this study is that the crossover
boundary between the two algorithms was characterized on
i.i.d.\ synthetic inputs and one real dataset.
A natural direction for future work is a controlled experiment
that varies spatial autocorrelation independently of density,
which would quantify precisely how much structure is needed for
\freqb{} to overtake \textsc{dp\_lin\_1arr} at low density. More broadly, this work illustrates that reducing algorithmic
work can outweigh reduced execution efficiency, and that
hardware performance counters provide useful insight into
the algorithmic mechanisms underlying observed speedups.


\bibliographystyle{plain}
\bibliography{references}

\clearpage
\appendix

\section{AMD Cross-Architecture Figures}
\label{app:amd-causal}

\begin{figure}[h]
  \centering
  \includegraphics[width=\linewidth]{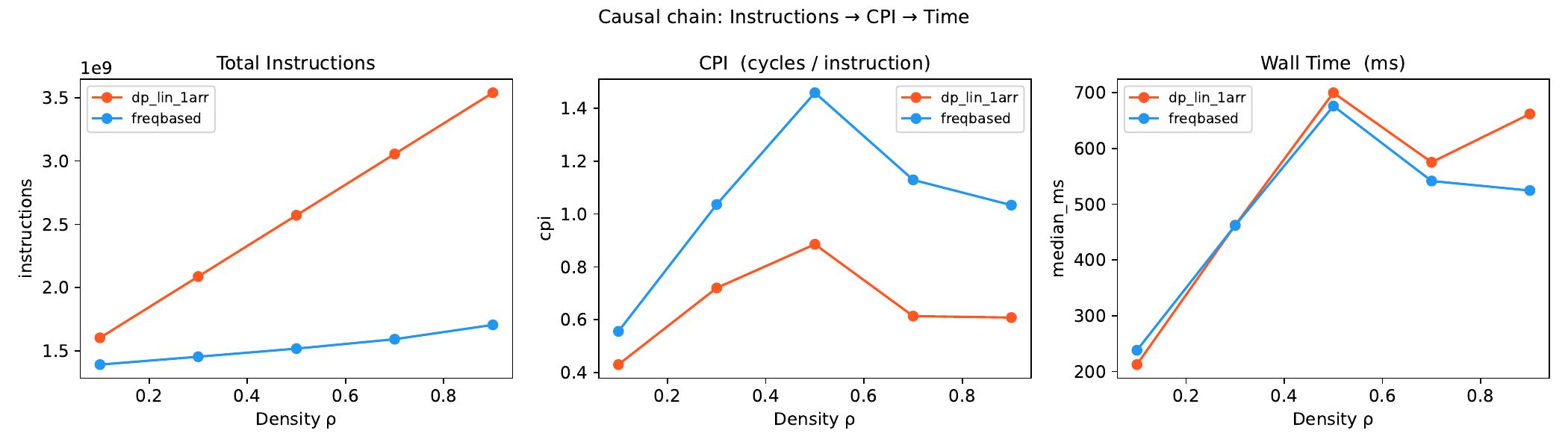}
  \caption{Causal decomposition on AMD Ryzen~5 7535HS: total
           instructions, CPI, and wall time versus density $\rho$.}
  \label{fig:amd-causal-chain}
\end{figure}

\begin{figure}[h]
  \centering
  \includegraphics[width=\linewidth]{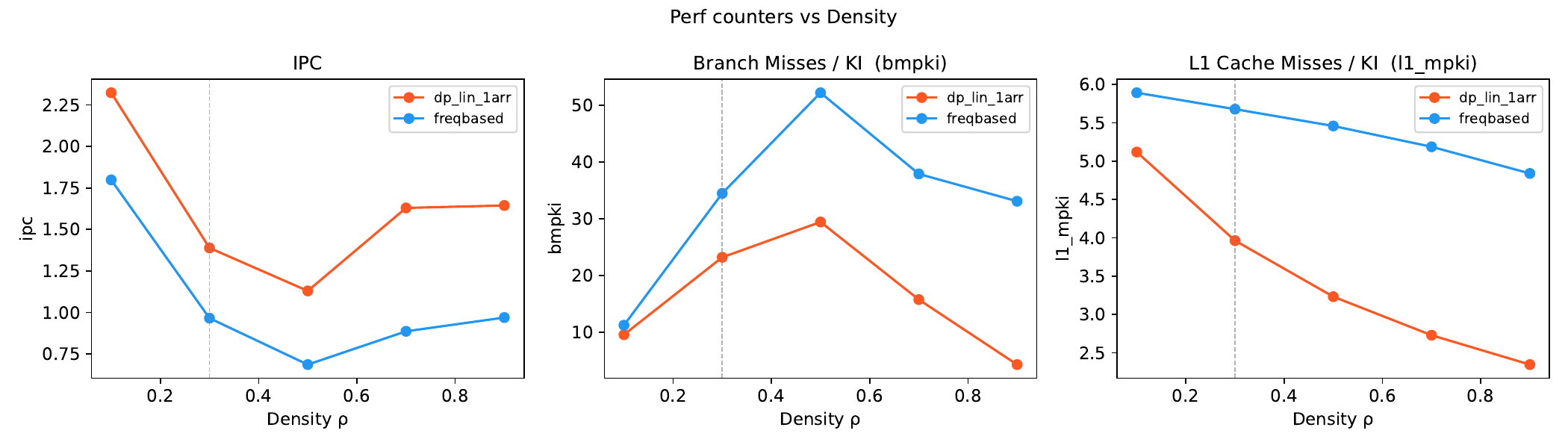}
  \caption{IPC, \texttt{bmpki}, and \texttt{l1\_mpki} versus
           density $\rho$ on AMD Ryzen~5 7535HS (\texttt{znver3}).}
  \label{fig:amd-counters-density}
\end{figure}

\begin{figure}[h]
  \centering
  \includegraphics[width=\linewidth]{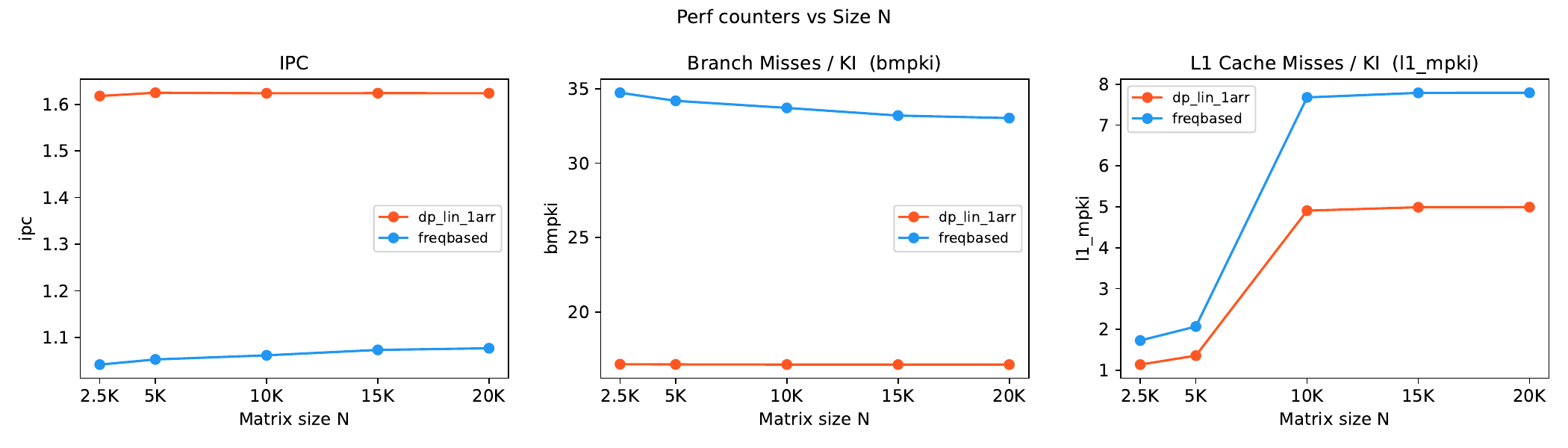}
  \caption{IPC, \texttt{bmpki}, and \texttt{l1\_mpki} versus
           matrix size $N$ on AMD Ryzen~5 7535HS (\texttt{znver3}).}
  \label{fig:amd-counters-N}
\end{figure}

Figures~\ref{fig:amd-causal-chain}--\ref{fig:amd-counters-N} show
the AMD Ryzen~5 7535HS counterparts to
Figures~\ref{fig:causal-chain}, \ref{fig:counters-density},
and~\ref{fig:counters-N} in the main text.
All qualitative patterns are consistent with the Intel results.

\section{Reproducibility}
\label{app:reproducibility}

The complete reproducibility package is available at:
\begin{center}
\url{https://www.dropbox.com/scl/fi/7mikg9qj5znusf1uewkjr/ALENEX2027.zip?rlkey=jf7wog30smi3faggtxxc4hicd&st=qhru831a&dl=0}
\end{center}
The package includes all source code, result CSVs, analysis notebooks,
generated figures, raw INRIA ground-truth masks, and derived binary
matrices, together with a captured system configuration from
\texttt{capture\_env.sh}.

\paragraph{Requirements.}
Linux (perf counters require \texttt{perf\_event\_open}), g++ with
C++17 support, and Python~3 with \texttt{numpy}, \texttt{pillow},
\texttt{pandas}, \texttt{matplotlib}, and \texttt{seaborn}.
Hardware performance counters require:
\begin{verbatim}
sudo sysctl -w kernel.perf_event_paranoid=0
\end{verbatim}

\paragraph{Three entry points.}
The pre-computed result CSVs ship with the package, giving three
starting points depending on how much needs to be re-executed.

\textit{Figures only.}
Run \texttt{plots.py} and \texttt{inriaDA.ipynb} directly against
the shipped CSVs (\texttt{results\_amd.csv},
\texttt{results\_intel.csv}, \texttt{results\_real\_inria.csv})
to regenerate all figures without re-running any benchmark.

\textit{Benchmarks.}
Recompile and re-run the synthetic and real benchmarks to regenerate
the CSVs:
\begin{verbatim}
g++ -O2 -march=native -DNDEBUG -flto \
    msbench_vv.cpp -o msbench_vv
sudo ./msbench_vv --label intel_g3 \
    --no-turbo --affinity 2 > results_intel.csv
sudo bash run_real.sh
\end{verbatim}
The synthetic benchmark executes a correctness gate of 150{,}000
brute-force comparisons before any measurement and aborts on any
mismatch; a completed run is therefore self-validating.

\textit{Full pipeline.}
To regenerate the binary matrices from the raw INRIA masks:
\begin{verbatim}
rm -rf real_matrices/inria/
sudo bash run_real.sh
\end{verbatim}
\texttt{prepare\_datasets.py} reproduces the binary matrices
byte-for-byte from the \texttt{.tif} masks using load, grayscale
conversion, and threshold at 127.
\end{document}